\documentclass[a4paper,fleqn,usenatbib]{mnras}

\usepackage{mathptmx}
\usepackage{aecompl}
\usepackage{graphicx}
\usepackage{times}
\usepackage{booktabs}
\usepackage{float}
\usepackage[english]{babel}
\usepackage{color}
\usepackage{amsmath}
\usepackage{chngcntr}

\usepackage[T1]{fontenc}
\usepackage{ae,aecompl}

\usepackage{graphicx}	
\usepackage{amsmath}	
\usepackage{amssymb}	

\title[X-ray background and 21~cm signal]
{X-ray background and its correlation with the 21~cm signal}

\author[Ma et al.]{
Q. Ma$^{1,2,3,4}$,\thanks{E-mail: maqb@mpa-garching.mpg.de}
B. Ciardi$^{2}$,
M. B. Eide$^{2}$,
K. Helgason$^{2,5}$
\\
$^1$ Purple Mountain Observatory, Chinese Academy of Sciences, Nanjing 210008, China\\
$^2$ Max-Planck-Institut f\"ur Astrophysik, Karl-Schwarzschild-Stra\ss e 1, D-85748 Garching bei M\"unchen, Germany\\
$^3$ Guizhou Provincial Key Laboratory of Radio Astronomy and Data Processing, Guizhou Normal University, Guiyang 550001, China\\
$^4$ University of Chinese Academy of Sciences, Beijing 100049, China\\
$^5$ Centre for Astrophysics and Cosmology, University of Iceland, Dunhagi 5, 107 Reykjav\'ik, Iceland 
}

\date{Accepted XXX. Received YYY; in original form ZZZ}

\pubyear{2017}

\begin{document}
\label{firstpage}
\pagerange{\pageref{firstpage}--\pageref{lastpage}}
\maketitle

\begin{abstract}
We use high resolution hydrodynamical simulations to study the contribution to the X-ray background from high-$z$ energetic sources, such as X-ray binaries, accreting nuclear black holes and shock heated interstellar medium. 
Adopting the model discussed in \citet{Eide2018}, we find that these X-ray sources during the Epoch of Reionization (EoR) contribute less than a few percent of the unresolved X-ray background.
The same sources contribute to less than $\sim$2\% of the measured angular power spectrum of the fluctuations of the X-ray background.
The outputs of radiative transfer simulations modeling the EoR are used to evaluate the cross-correlations of X-ray background with the 21~cm signal from neutral hydrogen.
Such correlation could be used to confirm the origin of the 21 cm signal, as well as give information on the properties of the X-ray sources during the EoR.
We find that the correlations are positive during the early stages of reionization when most of the hydrogen is neutral, while they become negative when the intergalactic medium gets highly ionized, with the transition from positive to negative depending on both the X-ray model and the scale under consideration.
With {\tt SKA} as the reference instrument for the 21~cm experiment, the predicted S/N for such correlations is $<1$ if the corresponding X-ray survey is only able to resolve and remove X-ray sources with observed flux $>10^{-15}\,\rm erg\, cm^{-2} \, s^{-1}$, while the cumulative S/N from $l=1000$ to $10^{4}$ at $x_{\rm HI}=0.5$ is $\sim 5$ if sources with observed flux $>10^{-17}\,\rm erg\, cm^{-2} \, s^{-1}$ are detected.

\end{abstract}



%
\section{INTRODUCTION}
\label{sec:intro}
The combined contribution of radiation from sources at all redshifts makes up the background radiation permeating our universe.
The X-ray component of such background includes input from a variety of high-energy sources  \cite[e.g.][]{Brandt2005,Helgason2014,Cappelluti2017}, such as X-ray binaries (XRBs) \citep{Fragos2013a}, active galactic nuclei (AGNs) \citep{Comastri1995}, and thermal bremsstrahlung from hot gas \citep{Cappelluti2012}. 

Deep X-ray surveys \citep{Lehmer2012,Luo2017} have resolved most of the cosmic X-ray background (CXB) as point sources which are dominated by AGNs, while $\sim 10-30\%$ of the flux remains unresolved (see e.g. \citealt{Lehmer2012, Moretti2012, Cappelluti2017}). More specifically,
the $4 \rm Ms$ {\tt Chandra} Deep Field-South (CDF-S) (\citealt{Lehmer2012}) showed that this component amounts to
24.3\% of the total flux in the soft (0.5-2)~keV band (i.e. $8.15 \pm 0.58 \times 10^{-12}\,\rm erg\, cm^{-2} \, s^{-1}\, deg^{-2}$) and 17.6\% in the hard (2-8)~keV band (i.e. $1.73\pm 0.23 \times 10^{-11}\,\rm erg\, cm^{-2} \, s^{-1}\, deg^{-2}$), 
which is composed by very faint sources that lay below the sensitivity of the instrument. 
Although these faint sources can not be singly resolved, the imprint of their angular fluctuations on the CXB could provide some information on their properties \citep{Yamamoto1998,Sliwa2001,Kolodzig2017,Kolodzig2018}. 
Specifically, measurements of cross-correlation with the near infrared background indicate that such fluctuations should partially come from the high-$z$ universe \citep{Cappelluti2012,Helgason2014,Fernandez2014,Mitchell2016}. 

It has been suggested that this high-$z$ component could originate from accretion powered sources during the epoch of reionization (EoR), such as XRBs \citep{Fragos2013a,Fragos2013b,Xu2016}, and/or AGNs \citep{Dijkstra2004,Salvaterra2005,Salvaterra2007}.
Although they are also expected to contribute to the CXB,
it is not possible to separate such components directly from the CXB measurements due to the lack of redshift information.
Since these sources are usually hosted in very dense regions, they are expected to anti-correlate with the 21~cm signal originating from the neutral hydrogen found further away from the production sites of ionizing photons \citep{Shan2009,Liang2016}.
Exploiting the redshift information offered by the 21~cm signal, such cross-correlation could then be used to give information on the properties of the X-ray sources during the EoR, as well as confirm the origin of the 21~cm signal.
It should be noted that the high-$z$ component of the X-ray background is mainly contributed by the hard X-ray radiation that can easily escape from the hosts and travel large distances, while the soft X-ray photons produced by the same sources are expected to interact with the neutral hydrogen and helium and impact the physical properties of the intergalactic medium, and thus the associated 21~cm signal \citep{Mesinger2013,Christian2013,Fialkov2014,Fialkov2017,Eide2018}.

In this paper, we will evaluate the contribution to the CXB from high-$z$ energetic sources such as XRBs, accreting nuclear black holes and shock heated interstellar medium. 
The properties of the sources are retrieved from the hydrodynamic simulation MassiveBlack-II \citep{Khandai2015}. 
We also employ the 3D multi-frequency radiative transfer code CRASH (\citealt{Ciardi2001,Maselli2009}; Graziani, Ciardi \& Glatzle sub.) to follow the reionization history of hydrogen and helium (\citealt{Eide2018} and Eide et al. in prep). 
These simulations are used to evaluate the 21~cm signal and the cross-correlation with the CXB for several source models.
As mentioned above, such correlations were previously investigated by \cite{Shan2009} by means of two semi-analytic models for reionization (stars dominated or quasars dominated), and by \cite{Liang2016} using semi-numerical 21CMFAST simulations which included three cases of X-ray emitting efficiency. Both works 
concluded that a cross-correlation between X-ray background and 21 cm signal exists, although it is difficult to measure.
Ours is the first work that attacks the problem employing a combination of hydrodynamic and radiative transfer simulations.

The rest of paper is organized as follows.
Simulations and X-ray models adopted are described in section~\ref{sec:simu};
the results are presented in section~\ref{sec:res}, while the conclusions are summarized in section~\ref{sec:dis}.
The cosmological parameters adopted are from WMAP7 \citep{Komatsu2011} within a  $\Lambda$CDM cosmology, i.e. $\sigma_{8} = 0.816$, $n_{s} = 0.968$, $\Omega_{\Lambda} = 0.725$,
$\Omega_{m} = 0.275$, $\Omega_{b} = 0.046$ and $h = 0.701$.
%
\section{SIMULATIONS}
\label{sec:simu}
The simulations used to compute X-ray background and 21 cm signal are those described in \citet[][ hereafter Eide18a]{Eide2018} and Eide et al. (in prep., hereafter Eide18b). Here we outline their main characteristics and refer the reader to the original papers for more details. 

The gas and galaxy distribution adopted, together with their properties, are obtained from the high resolution cosmological hydrodynamical simulation MBII presented in \cite{Khandai2015}, with box length $100\,h^{-1} \rm cMpc$ and $2 \times 1792^{3}$ particles, i.e. a dark matter and gas particle mass of $m_{\rm DM} = 1.1 \times 10^7\,h^{-1}$ M$_{\odot}$ and $m_{\rm gas} = 2.2 \times 10^{6}\,h^{-1}$ M$_{\odot}$, respectively.
The simulation tracks stellar populations, galaxies, accreting and dormant black holes (growing from seeds of 10$^5$~M$_\odot$ by merging with other black holes and by accreting gas at a maximum of twice the Eddington rate), as well as their properties, such as mass, age, metallicity, accretion rate, and star formation rate (SFR).
The baryonic physics and feedback effects of the sources are also accounted for.
We refer the reader to \cite{Khandai2015} for more details about the hydrodynamical simulations.
Here we just note that the parameters involved in the calculations have been tuned to reproduce low redshift observations, such as the cosmic SFR, the galaxy stellar mass function and the quasar (QSO) bolometric luminosity function. 

The outputs of the simulations have been mapped onto $N_c=256^3$ grids and post-processed with the Monte Carlo 3D radiative transfer code CRASH \citep{Ciardi2001,Maselli2003,Maselli2009,Graziani2013} to follow the redshift evolution of the ionization and temperature state of the intergalactic medium (IGM), i.e. the reionization process, as determined by different source types (and their combination) emitting in the energy range 13.6~eV-2~keV.
In addition to stellar type sources that emit the bulk of their radiation in the UV band, Eide18a,b considered the contribution to the production of more energetic photons from:  
\begin{enumerate}
  \item {\it X-ray binaries (XRBs)}. These include low mass XRBs (LMXBs) and high mass XRBs (HMXBs). The corresponding total luminosity in a galaxy is given by the contribution from both populations: $L^{\rm XRB} = L^{\rm HMXB}+ L^{\rm LMXB}$.
  The luminosity of HMXBs relates to the SFR of the galaxy and the metallicity ($Z$) of the residing stars as \citep{Fragos2013b, Madau2017}:
  \begin{equation}
  {\rm log}(L^{\rm HMXB}/{\rm SFR})=\displaystyle\sum_{i} \beta_{i} Z^{i},
  \end{equation}
  while the luminosity of LMXBs is determined by the mass ($M$) and age ($t$) of the stars:
  \begin{equation}
  {\rm log}(L^{\rm LMXB}/M)=\displaystyle\sum_{i} \gamma_{i} ({\rm log}{t})^{i},
  \end{equation}
  where $\beta_{i}$ and $\gamma_{i}$ denote the best-fit coefficients from \cite{Madau2017}.

  The spectra of the XRB sources, which are only mildly dependent on the redshift, are taken from \cite{Fragos2013b}.

  \item {\it Supernova heated interstellar medium (ISM)}. For each galaxy, the luminosity of the ISM in the energy band (0.3-10)~keV is evaluated as \citep{Mineo2012_ism}:
      \begin{equation}
      L^{\rm ISM}=(7.3\pm1.3)\times10^{39} \, {\rm erg\, s^{-1}\, {\rm M}_{\odot}^{-1}\, yr} \times {\rm SFR},
      \end{equation}
  where ${\rm SFR}$ is the total SFR in the galaxy. The luminosity is then rescaled to the frequency range of our interest.

  We assume that the ISM spectrum is thermal bremsstrahlung and constant in redshift \citep{Pacucci2014}:
  \begin{equation}
  S^{\rm ISM}(\nu)= \begin{cases}
                      C       & \quad \text{if}\,\, h_{\rm P}\nu\le kT_{\rm ISM}\\
                      C(h_{\rm P}\nu/kT_{\rm ISM})^{-3}  & \quad \text{if}\,\, h_{\rm P}\nu \ge kT_{\rm ISM}
  \end{cases}
  \end{equation}
  where $C$ is the normalization constant, $h_{\rm P}$ denotes the Planck constant, and $T_{\rm ISM}\sim 10^6$~K is the temperature of the heated ISM.

  \item {\it Accreting nuclear black holes (BHs)}. For each source we assume a bolometric luminosity \citep{Shakura1973}:
      \begin{equation}
      L^{\rm BH}=\eta \dot{M}_{\rm BH}c^{2},
      \end{equation}
      where $\eta=0.1$ is the efficiency parameter used in \cite{Khandai2015}, and $\dot{M}_{\rm BH}$ is the accretion rate of the black hole computed in the simulations.

      The spectrum is based on observational data by \cite{Krawczyk2013}, which is modeled as a power law with index $\alpha=-1$ at $h_{\rm P}\nu>200$~eV and has no evolution with redshift.
\end{enumerate}

The suite of simulations used in this paper is composed of four runs from Eide18a and Eide18b, including:  
stars and XRBs (XRB model), stars and ISM (ISM model), stars and BHs (BH model), and all of sources combined (referred to as ``Total'' model). 
We refer the reader to the original papers for more details on the simulations.

It should be noted that in the RT simulations only photons with energies below 2~keV are included as the mean free path of the ones with higher energy is larger than the box size. 
Those soft X-ray photons do not contribute to the CXB, but  Eide18a showed that they have an impact on the heating and ionization of the IGM, although full hydrogen ionization is driven by the UV radiation from stars. 

%
\section{RESULTS}
\label{sec:res}
%
%

\begin{figure}
\centering
\includegraphics[width=0.9\linewidth]{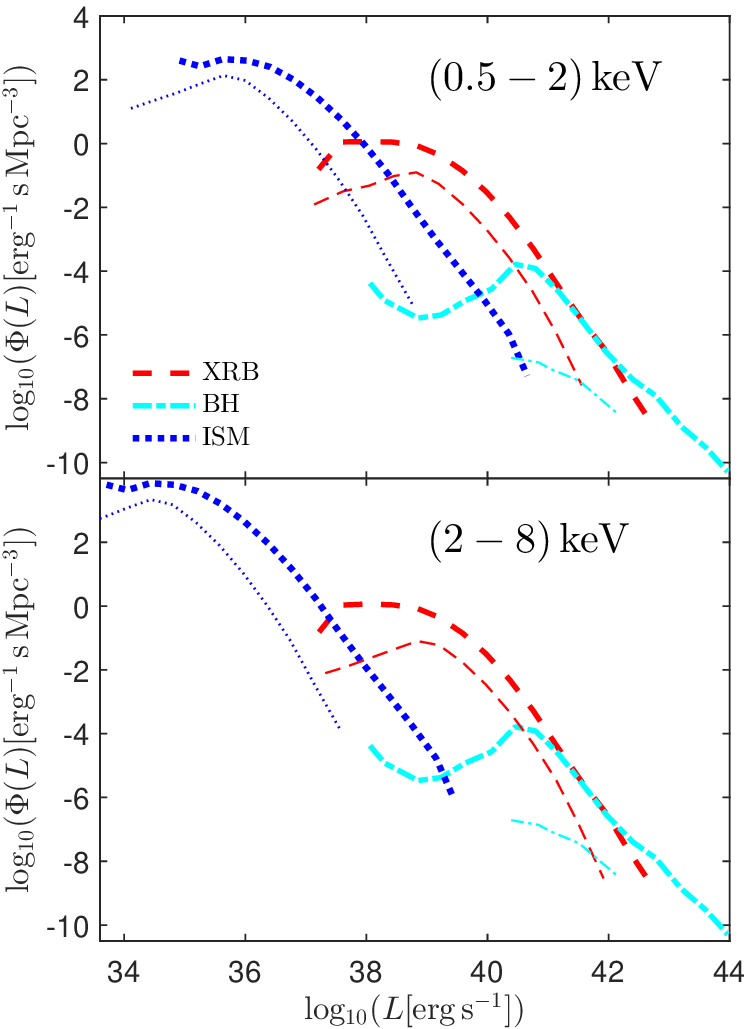}
\caption{Intrinsic luminosity function in the observer frame frequency bands (0.5-2)~keV (top panel) and (2-8)~keV (bottom) as produced by our population of XRBs (red dashed lines), BHs (cyan dash-dotted) and ISM (blue dotted). Thick and thin lines refer to $z=5$ and 10, respectively. 
}
\label{lum_func}
\end{figure}
Fig.~\ref{lum_func} shows the intrinsic luminosity function of X-ray radiation as contributed by our population of XRBs, BHs and ISM in the observer frame frequency bands (0.5-2)~keV and (2-8)~keV. The X-ray luminosity associated to the ISM is much lower than that of the others, while the BHs reach the highest luminosities, despite having the lowest number density.
Due to its softer X-ray spectrum (i.e. $\propto \nu^{-3}$), the luminosity of the ISM is smaller in the (2-8)~keV band than that in the (0.5-2)~keV band, while the luminosities of XRBs and BHs are similar in the two bands.

As single X-ray sources at high redshift are too faint to be resolved as point sources (e.g. the observed flux for a source with $L=10^{44}\, \rm erg\,s^{-1}$ at $z=5$ is $3.6 \times 10^{-16} \,\rm erg\,cm^{-2}\,s^{-1}$), most of them only contribute to the unresolved CXB\footnote{The brightest sources could be resolved in very deep observations with small fields of view \cite[e.g.][]{Lehmer2012,Luo2017}, but it would be difficult to detect them in large scale surveys \cite[e.g.][]{Cappelluti2012, Cappelluti2013,Kolodzig2017}.}.

\subsection{Global X-ray Flux}
%
%
The global X-ray flux measured on earth from a source type $s$ (XRB, BH or ISM) at redshift larger than $z$ is expressed as:
\begin{equation}
F^{s}(>z)=\frac{c}{4\pi}\int_{>z} \frac{{\rm d}z'}{H(z')} \frac{\langle L^{s} \rangle}{(1+z')^2} \int_{\nu_{\rm min}(1+z')}^{\nu_{\rm max}(1+z')} {\rm d}\nu S^{s}(\nu,z') e^{-\tau(\nu,z')},
\end{equation}
where $c$ is the speed of light, $H(z')=H_{0}\sqrt{\Omega_{m}(1+z')^{3}+\Omega_{\Lambda}}$ is the cosmic expansion rate at $z'$, $\langle L^{s} \rangle$ is the averaged luminosity per unit volume, and $\tau(\nu,z')$ is the optical depth from $z'$ to the observer for photons with frequency $\nu$:
\begin{align}
    \tau(\nu,z') = \int_{0}^{z'} \frac{c  {\rm d}z''}{(1+z'')H(z'')}  \sum_{m} \sigma_{m}(\nu') \langle n_{m} \rangle (z''),
\end{align}
where $\nu'=\nu (1+z'')/(1+z')$, $\sigma_{m}$ is the cross section to X-ray photons for species $m$=HI, HeI and HeII \citep{Verner1996}, and $\langle n_{m} \rangle$ is the corresponding volume averaged number density as obtained in Eide18a and Eide18b. 
More specifically, $\langle n_m \rangle = \sum_c n_m^c / N_c$, where $n_m^c$ is the number density of species $m$ in cell $c$ and it is evaluated from the gas number density and the ionization fractions of hydrogen and helium.
As the simulations have been run only until full hydrogen reionization has been reached, $z_{\rm r}\sim6$, we assume that at $z<z_{\rm r}$ hydrogen remains fully ionized, while for helium we assume that $x_{\rm HeII}(3<z<z_{\rm r})=x_{\rm HeII}(z_{\rm r})$, $x_{\rm HeII}(z\le 3)=0$, $x_{\rm HeIII} (3<z<z_{\rm r})=x_{\rm HeIII}(z_{\rm r})$, and $x_{\rm HeIII}(z\le 3)=1$.
As expected, the optical depth to the energetic photons of interest here is extremely low (e.g. $\tau<0.0003$ for photons $> 0.5$~keV in the observer frame at $z<15$), thus it can be safely neglected. 

\begin{figure}
\centering
\includegraphics[width=0.9\linewidth]{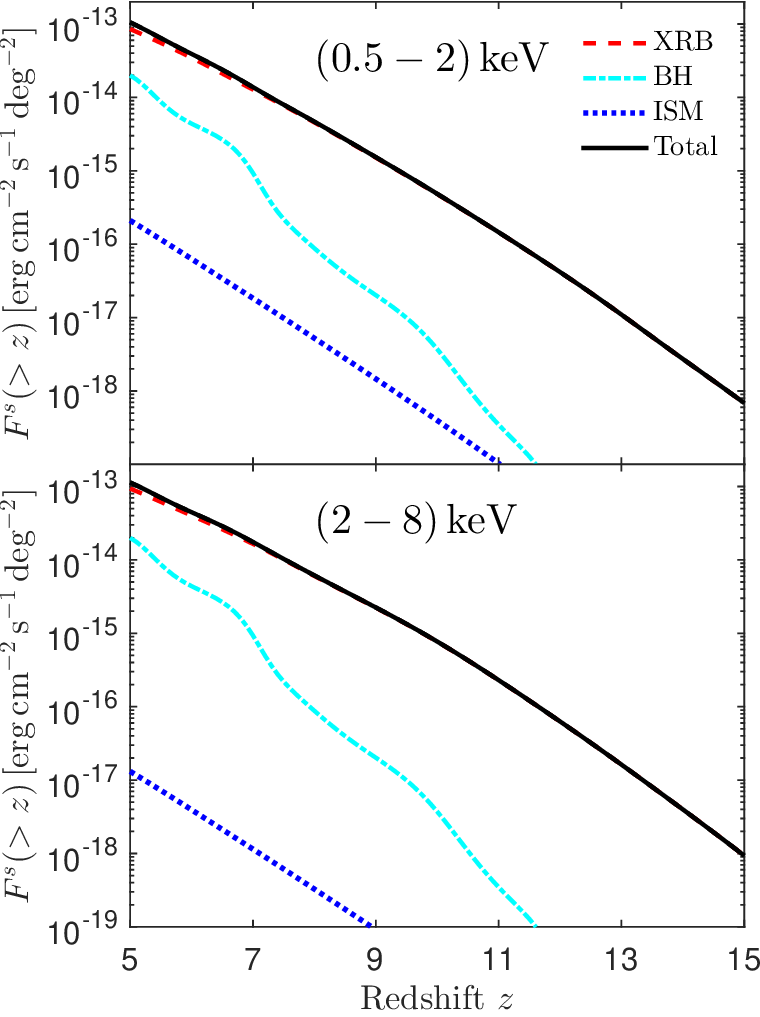}
\caption{Integrated X-ray flux at $>z$ for source model XRB (dashed red lines), BH (dot-dashed cyan), ISM (dotted blue) and Total (solid black). The upper and lower panel refers to the energy band (0.5-2)~keV and (2-8)~keV, respectively. 
}
\label{flux_df_dllz}
\end{figure}
Fig.~\ref{flux_df_dllz} presents the predicted global X-ray flux in the observed (0.5-2)~keV and (2-8)~keV bands.
For all models, the flux increases quickly with decreasing redshift, due to the fast growing of structures and the associated relevant physical properties, e.g. stellar mass, SFR, accreting black holes (see Eide2018a,b for further discussions).
The XRB and BH models display X-ray fluxes very similar in the soft and hard bands due to their spectral index (for both it is $\approx -1$), while the ISM model shows a lower flux in the hard than in the soft band as a consequence of its softer spectrum (spectral index $-3$).
As the flux in the Total model is dominated at all redshifts by the XRBs contribution, its behavior is similar to that of the XRB model.

In the soft (0.5-2)~keV band the flux from XRBs, BHs and ISM at $z>5$ is $8.18 \times 10^{-14}$, $1.99 \times 10^{-14}$,  and $2.04 \times 10^{-16} \,\rm erg\, cm^{-2}\, s^{-1}\, deg^{-2}$, respectively, corresponding to 80.3\%, 19.5\% and 0.2\% of the flux in the Total model, which is $1.02 \times 10^{-13} \,\rm erg\, cm^{-2}\, s^{-1}\, deg^{-2}$. 
In comparison, the unresolved CXB in the same band measured by \cite{Cappelluti2017} is $(2.90 \pm 0.16) \times 10^{-12} \,\rm erg\, cm^{-2}\, s^{-1}\, deg^{-2}$, while
the unresolved X-ray flux inferred by \cite{Lehmer2012} is
$(1.98 \pm 0.35) \times 10^{-12} \,\rm erg\, cm^{-2}\, s^{-1}\, deg^{-2}$, i.e. our predicted X-ray flux from the EoR is less than a few percent of the unresolved CXB expected from observations.

Similarly, in the hard (2-8)~keV band the flux from XRBs, BHs and ISM at $z>5$ is $8.94 \times 10^{-14}$, $ 1.99\times 10^{-14}$, and $1.28\times 10^{-17} \,\rm erg\, cm^{-2}\, s^{-1}\, deg^{-2}$, respectively, corresponding to 81.8\%, 18.2\% and $\sim0.01\%$ of the flux in the Total model, which is $1.09 \times 10^{-13} \,\rm erg\, cm^{-2}\, s^{-1}\, deg^{-2}$.
The unresolved X-ray flux in the same band inferred by \cite{Lehmer2012} is $(3.05 \pm 2.25)\times 10^{-12} \,\rm erg\, cm^{-2}\, s^{-1}\, deg^{-2}$, while in the similar band (2-10)~keV the unresolved CXB measured by \cite{Cappelluti2017} is $(6.47 \pm 0.82)\times 10^{-12} \,\rm erg\, cm^{-2}\, s^{-1}\, deg^{-2}$. 

In the following we will present results only in the soft band, as those in the hard band are similar.

\subsection{Auto Power Spectra}
While it is difficult to measure directly the X-ray sources during the EoR due to their faint luminosities, it might be possible to probe them through the angular distribution of the CXB.

The angular auto power spectrum of the X-ray flux relates to the 3D power spectrum of X-ray sources, $P_{\rm X}^{s}(k,z)$, through Limber's approximation\footnote{ Limber's approximation is consistent with the exact power spectrum at $l>10$, while it breaks at lower $l$ \cite[see e.g. ][]{Simon2007,Loverde2008}.} \citep{Limber1953}: 
\begin{equation}
\label{eq:c_x}
C_{\rm X}^{s}(l)=\int_{z>5} {\rm d}z \frac{H}{c d_{\rm com}^2} \Psi_{\rm X}^{s}(z)^{2}  P_{\rm X}^{s}(k=\frac{l}{d_{\rm com}},z),
\end{equation}
where $d_{\rm com}$ is the comoving distance from 0 to $z$, and
\begin{equation}
\Psi_{\rm X}^{s}(z)=\frac{c}{4\pi}\frac{1}{H(z)(1+z)^2} \int_{\nu_{\rm min}(1+z)}^{\nu_{\rm max}(1+z)}{\rm d}\nu S^{s}(\nu,z) e^{-\tau(\nu,z)}.
\end{equation}
$P_{\rm X}^{s}(k,z)$ is computed from the luminosity density of the X-ray sources contained in the snapshot of the simulation at $z$. 

\begin{figure}
\centering
\includegraphics[width=0.9\linewidth]{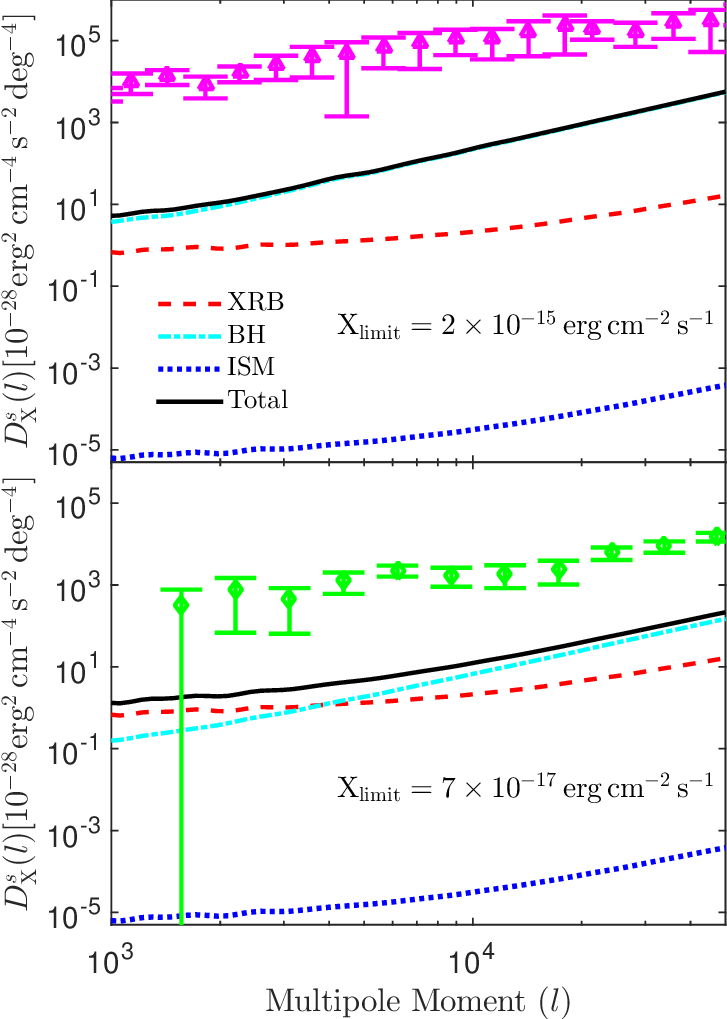}
\caption{Angular power spectrum of the X-ray flux in the (0.5-2)~keV band from $z>5$. The lines refer to the XRB (dashed red lines), BH (dash-dotted cyan), ISM (dotted blue) and Total (solid black) model. The top panel is the power spectra obtained when bright sources with observed flux $\ge 2\times 10^{-15} \rm erg\, cm^{-2}\, s^{-1}$ are removed, while the bottom panel removes bright sources with observed flux $\ge 7 \times 10^{-17} \rm erg\, cm^{-2}\, s^{-1}$. 
The green diamond data points with error bars in the bottom panel are from \citet{Cappelluti2013} obtained with a flux limit of $\sim 7\times 10^{-17} \rm erg\, cm^{-2}\, s^{-1}$, while the magenta triangle ones in the top panel are from \citet{Kolodzig2017} obtained with a flux limit of $\sim 2\times 10^{-15} \rm erg\, cm^{-2}\, s^{-1}$.
}
\label{ps_xray}
\end{figure}

The angular power spectra $D_{\rm X}^{s}(l)=l(l+1)C_{\rm X}^{s}(l)/2\pi$ from the four models are shown in Fig.~\ref{ps_xray}. 
The power spectrum of the ISM model displays an amplitude much lower than the others, consistently with its very low global flux showed in Fig.~\ref{flux_df_dllz}. 
Due to the rarity of BHs on the small scales, the power spectrum of the BH model approaches a white noise distribution with increasing $l$ (i.e. $D_{\rm X}^{\rm BH} \propto l^{2}$) and its amplitude is higher than the ones in the other models, although the global flux produced by BHs is lower than the one originating from XRBs (see Fig.~\ref{flux_df_dllz}). For the same reason, the power spectrum in the Total model is dominated by BHs.
This can be clearly seen in the top panel of Fig.~\ref{ps_xray}, where the power spectra have been calculated including all sources with an observed flux $<2 \times 10^{-15} \rm erg\, cm^{-2}\, s^{-1}$, to be compared to results from the XBOOTES survey (with a surface area of $\sim 9\,\rm deg^{2}$; \citealt{Kolodzig2017}) that detected and removed sources with observed flux $> 2\times 10^{-15} \rm erg\, cm^{-2}\, s^{-1}$. 
Meanwhile, our predicted power spectra at all scales and in all models are much lower (below a few percent) than the currently measured one.

If sources could be detected and removed in deep X-ray surveys down to a lower observed flux, the contribution from BHs would be reduced, without substantially changing the spectra of XRBs and ISM as they are dominated by fainter sources. 
The bottom panel in Fig.~\ref{ps_xray} refers to the case in which sources with an observed flux $>7 \times 10^{-17} \rm erg\, cm^{-2}\, s^{-1}$ have been removed. Although at very small scales the power spectrum is still dominated by BHs, at $l<3000$ the Total model is now mainly determined by XRBs, and the impact of XRBs is expected to increase as more sources get removed.
The power spectrum in the Total model is $\lesssim 2\%$ of the one measured by \cite{Cappelluti2013}, who performed the same source subtraction in the deep {\tt Chandra ACIS-I AEGIS-XD} survey that covered approximately $0.1\,\rm deg^{2}$. 

\subsection{Cross-correlation with the 21~cm signal}
In Eide2018a and Eide2018b we have seen that the hard UV and soft X-ray photons emitted by the sources considered here have a substantial impact on the physical properties of the IGM, in particular its temperature and HeIII content. 
In a companion paper, we will investigate their effect on the 21~cm signal, while here we concentrate specifically on its correlation with the CXB. 
As mentioned above, the photons contributing to the CXB in the bands studied here have energies higher than those in Eide2018a and Eide2018b, i.e. they are not responsible for the formation and evolution of highly ionized regions. 
On the other hand, the 21~cm signal is expected to correlate negatively with the X-ray sources located within highly ionized regions, since the XRBs, BHs and hot ISM reside in the same galaxies hosting the stellar sources that drive IGM reionization. 
It should be noted that, while there is a one to one correlation between the sources of ionization and the CXB when this is dominated by XRBs, this is not strictly true for BHs as, despite still tracing highly ionized regions, there is not an active BH in each galaxy. 
In any case, such anti-correlation could be used to confirm the origin of the 21~cm signal, as well as give information on the properties of the X-ray sources during the EoR.

The 21~cm signal is typically described in terms of the differential brightness temperature, $\delta T_{\rm 21cm}$, defined as  \citep{Furlanetto2006}:
\begin{equation}
\label{eq:21cm}
 \delta T_{\rm 21cm} = 28{\rm mK} (1+\delta)x_{\rm HI}\left(1-\frac{T_{\rm CMB}}{T_{\rm s}}\right) \left(\frac{1+z}{10}\right)^{1/2},
\end{equation}
where $\delta$ denotes the gas matter overdensity, $T_{\rm CMB}=2.73 (1+z)$~K is the CMB temperature at $z$, and $T_{\rm s}$ is the spin temperature. 
$\delta T_{\rm 21cm}$ is calculated in each cell using the values of $\delta$, $x_{\rm HI}$ and the kinetic temperature of the gas, $T_k$, obtained in the simulations, and assuming that $T_{\rm s}=T_k$. 

The angular cross-power spectra of 21~cm brightness temperature and X-ray background is defined as $C_{\rm X-21cm}^{s}(z)\equiv \langle \widetilde{\delta T}_{\rm 21cm}  \widetilde{\delta F^{s}}^{\ast} \rangle$, where $\delta F^{s}$ denotes the fluctuation of the X-ray background, while $\widetilde{
\;}$ \, refers to the Fourier transfer.
With Limber's approximation, the cross-power spectra is expressed as:
\begin{equation}
C_{\rm X-21cm}^{s}(l,z)=\frac{H}{c d_{\rm com}^2} \Psi_{\rm X}^{s}(z) P_{\rm X-21cm}^{s}(k=\frac{l}{d_{\rm com}},z),
\end{equation}
where $P_{\rm X-21cm}^{s}$ is the 3D cross-power spectrum of X-ray sources and 21 cm signal computed from the output of the radiative transfer simulations.

\begin{figure}
\centering
\includegraphics[width=0.9\linewidth]{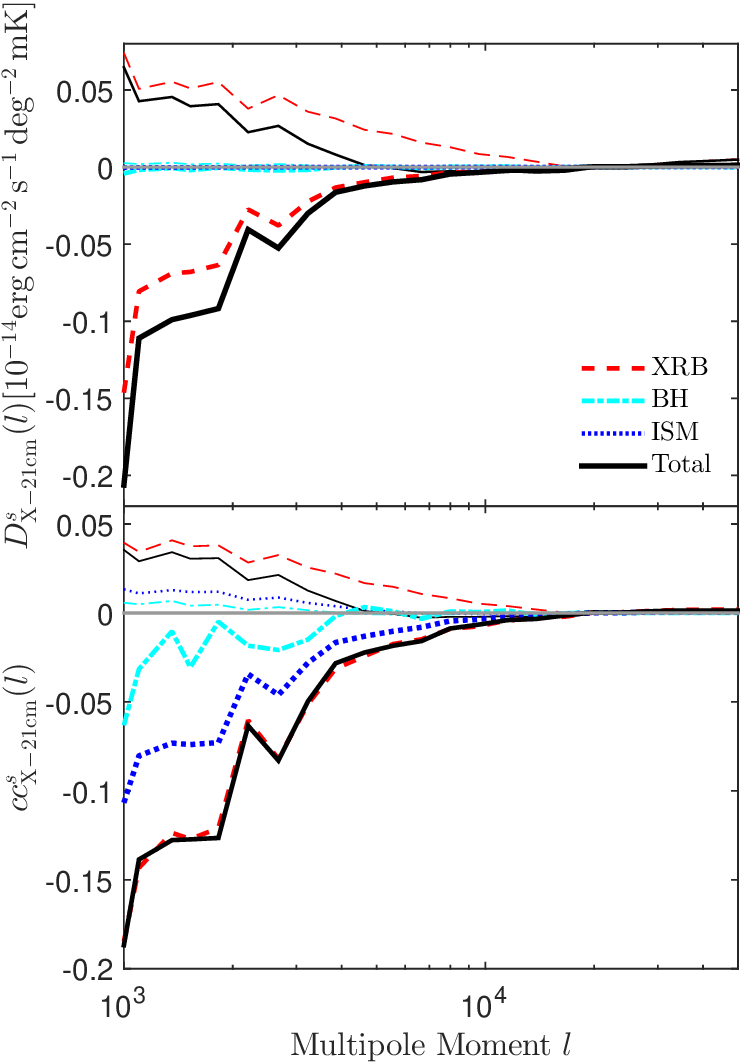}
\caption{Cross-power spectra (top) and correlation coefficients (bottom) between the X-ray background and the 21~cm brightness temperature at $z=7$ (thick lines), and 9 (thin) from the XRB (dashed red), BH (dash-dotted cyan), ISM (dotted blue) and Total (solid black) models.
Note that X-ray sources with an observed flux larger than $10^{-17} \rm erg\, cm^{-2}\, s^{-1}$ are removed.
The gray solid horizontal line denotes the zero of the y-axis. 
}
\label{ps_xray_21cm_c_vl}
\end{figure}
In the top panel of Fig.~\ref{ps_xray_21cm_c_vl} we show the angular cross-power spectra at $z=7$ and 9 for all simulations. 
At these redshifts the volume averaged HI fraction is $x_{\rm HI} \sim $ 0.24 and 0.94, respectively, for all models,
since the inclusion of more energetic sources has very little effect on the global hydrogen ionization fraction. 
It should be noted that here we only show a case in which X-ray sources with an observed flux larger than $10^{-17} \rm erg\, cm^{-2}\, s^{-1}$ have been removed.

In the bottom panel of Fig.~\ref{ps_xray_21cm_c_vl} we show the corresponding correlation coefficient factor, defined as:
\begin{equation}
cc_{\rm X-21cm}^{s}(l,z)=\frac{C_{\rm X-21cm}^{s}(l,z)}{\sqrt{C_{\rm X}^{s}(l)\times C_{\rm 21cm}^{s}(l,z)}},
\end{equation}
where $C_{\rm X}^{s}$ is the integral X-ray power spectra at $z>5$ (i.e. Eq.~\ref{eq:c_x}), and $C_{\rm 21cm}^{s}(l,z)$ is the angular power spectra of 21 cm brightness temperature at $z$.

For all models and at all redshifts, the correlations are significant only at $l<10^4$, i.e. physical scales $> 5.6$ cMpc ($k<1.13$ Mpc$^{-1}$) at $z=7$ and $> 6.0$ cMpc ($k<1.05$ Mpc$^{-1}$) at $z=9$.
At both redshifts, the cross-spectra in the ISM and BH models are much smaller than those in the XRB and Total models, consistently with their smaller global flux and angular power spectra (see Fig.~\ref{flux_df_dllz} and Fig.~\ref{ps_xray}).
At $z=9$, when most of the IGM is still neutral, the 21 cm brightness temperature is dominated by the contribution from the overdensity and the spin temperature (see Eq.~\ref{eq:21cm}).
As the sources of X-ray radiation mainly reside in overdense regions, the correlations are generally positive in all models, i.e. the X-ray radiation and the 21~cm signal during the early stage of reionization have a similar origin.
Due to the higher X-ray radiation and stronger heating, the positive signal in the Total model is weaker than that in the XRB model.
This behaviour is observed both in the power spectra and the correlation coefficients.
At $z=7$, when the IGM is highly ionized, a negative correlation is observed for all models, since the 21 cm brightness temperature in this case is more sensitive to the neutral hydrogen fraction which anti-correlates with the X-ray sources. Here the correlation is stronger than at $z=9$ because of the higher X-ray radiation at lower redshift.

\begin{figure}
\centering
\includegraphics[width=0.9\linewidth]{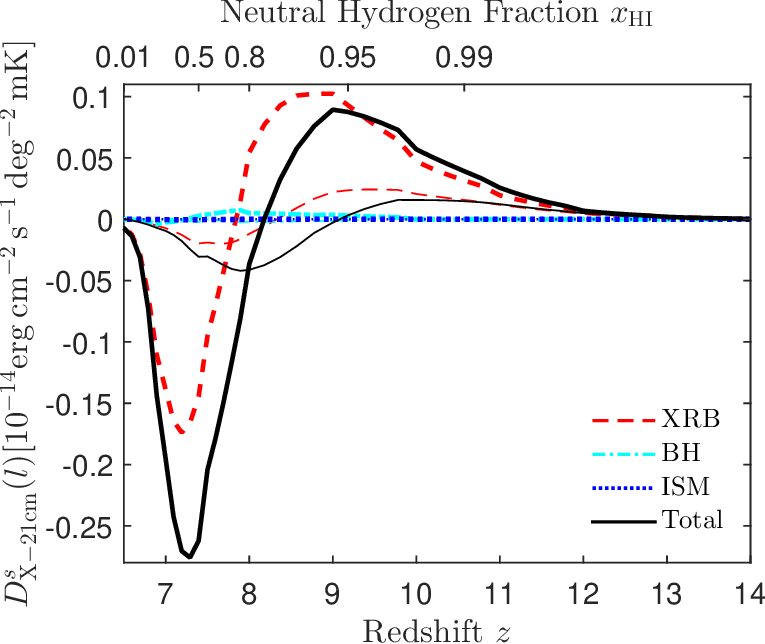}
\caption{Redshift evolution of the cross-power spectra between the X-ray background and the 21~cm  brightness temperature at $l=1000$ (thick lines) and 5000 (thin) for the XRB (dashed red), BH (dash-dotted cyan), ISM (dotted blue),  and Total (solid black) models. 
}
\label{ps_xray_21cm_c_vz}
\end{figure}
Fig.~\ref{ps_xray_21cm_c_vz} shows the redshift evolution of the cross-power spectra at $l=1000$ ($k \approx 0.11$ Mpc$^{-1}$ at $z=7$) and 5000 ($k \approx 0.56$ Mpc$^{-1}$ at $z=7$) for the four models.
As the amplitude of the cross-power spectra in the BH and ISM models is much lower than those in the XRB and Total models, we limit our discussion to the latter. 
The correlations are only significant at $z<12$, because of the much lower X-ray flux at higher $z$. 
At $z>9$, the cross-power spectra are positive at both $l=1000$ and 5000, while negative at $z<8$ where $x_{\rm HI} < 0.8$.
The transition from positive to negative depends on the models, as well as the multipole moment $l$.
In both models the transition happens earlier at $l=5000$ than at $l=1000$, since in a standard inside-out reionization scenario the smaller scales are ionized earlier.
As the X-ray heating is stronger in the Total than in the XRB model, this leads to the earlier transition of the Total model.

\subsection{Detectability}
The signal-to-noise ratio (S/N) for the cross-correlation power spectra at multiple $l$ can be estimated by \citep{dore2004}:
\begin{equation}
\label{eq:sn_cal}
    \left(\frac{S}{N}\right)^{2}=\frac{f_{\rm sky}(2l+1) l_{\rm bin} C_{\rm X-21cm}^{2}} {(C_{\rm X}+N_{\rm X}) (C_{\rm 21cm} + N_{\rm 21cm})+C_{\rm X-21cm}^{2}},
\end{equation}
where $f_{\rm sky}$ is the fraction of the sky covered by both the X-ray and 21~cm telescopes, $l_{\rm bin}$ is the bin width, $N_{\rm X}$ is the angular power spectrum of the X-ray foreground noise, and $N_{\rm 21cm}$ is the noise power spectrum of the 21~cm instrument.
We assume $f_{\rm sky}=0.0024$ (i.e. covered sky area $=100\,\rm deg^{2}$),
and $l_{\rm bin} \sim 0.46 l$ (i.e. $({\rm log}_{10} l)_{\rm bin}=0.2$). 

We adopt the X-ray foreground model by \cite{Helgason2014}, which includes X-ray radiation from  AGNs, hot gas and galaxies (mostly dominated by XRBs).
In this model, X-ray sources above a given flux limit $\rm X_{limit}$ are removed, assuming that they can be detected as point sources in deep X-ray surveys. 
For example, {\tt Chandra} is able to detect point sources with measured flux above $6.4\times10^{-18}\, \rm erg\, cm^{-2}\, s^{-1}$ in the soft band with an exposure time $\sim$ 7~Ms \citep{Luo2017}.
Here, we take three flux limits: $\rm X_{limit}=10^{-15}$, $10^{-16}$ and $10^{-17}\, \rm erg\, cm^{-2}\, s^{-1}$. 
Correspondingly, we compute $C_{\rm X}$ and $C_{\rm X-21cm}$ in Eq.~\ref{eq:sn_cal} by removing the sources brighter than those flux limits. 
While the instrument noise of X-ray facilities depends on the instrumental background and the exposure time, in the deep surveys it is expected to be much lower than the foreground contribution and/or to be removed in an efficient way \citep{Cappelluti2013,Kolodzig2017}, and thus we neglect it when estimating S/N.

We assume that the foreground contributions to the 21~cm signal can be accurately removed (see e.g. \citealt{Koopmans2015}).
The noise power spectrum for the 21~cm instrument can be expressed as \citep{Knox1995}:
\begin{equation}
N_{\rm 21cm}=[(1+z)/9.5)]^{2}\sigma_{\rm pix}^{2} \theta^{2} e^{l^2\theta^2/[8{\rm lg}(2)]},
\end{equation}
where $\sigma_{\rm pix}$ is the noise per pixel and $\theta$ is the full width at half maximum (FWHM) of the experimental beam. 
Using SKA\footnote{https://www.skatelescope.org} as our reference instrument, $\theta=1 \rm arcmin$ and $\sigma_{\rm pix}=1\,\rm mK$ at 150 MHz, corresponding to $\sim 1000$ hours of integration and 1 MHz of bandwidth \citep{Koopmans2015}. 

\begin{figure}
\centering
\includegraphics[width=0.9\linewidth]{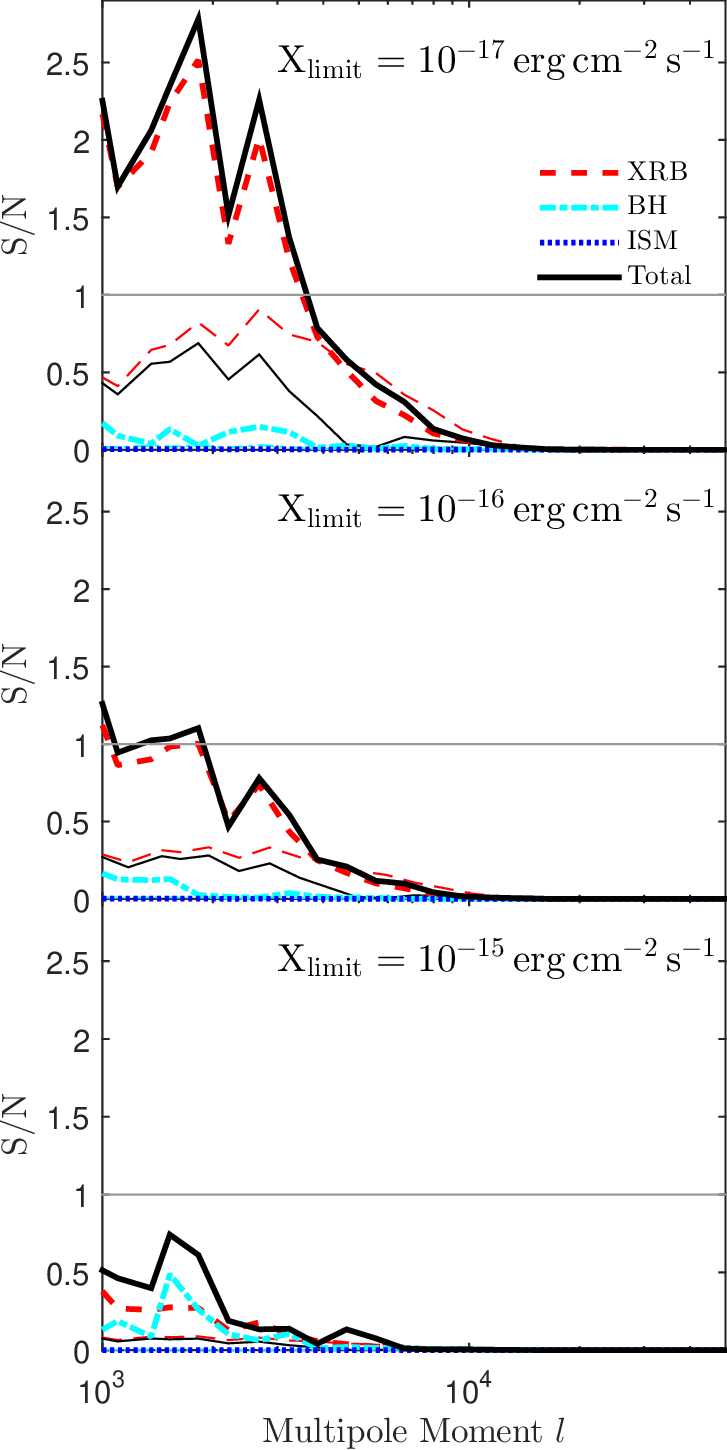}
\caption{S/N for X-ray flux limits of $10^{-17}$ (top panel), $10^{-16}$ (center) and $10^{-15}$ (bottom) $\, \rm erg\, cm^{-2}\, s^{-1}$  in model XRB (dashed red lines), BH (dash-dotted cyan), ISM (dotted blue)  and Total (solid black) at $z=7$ (thick lines) and $z=9$ (thin).
The gray solid horizontal line denotes S/N$=1$.
}
\label{sn_bin_vl}
\end{figure}
Fig.~\ref{sn_bin_vl} shows the predicted S/N for our four models and different X-ray flux limits.
For $\rm X_{limit}=10^{-15}\, \rm erg\, cm^{-2}\, s^{-1}$, S/N$<1$ for all models. 
When decreasing $\rm X_{limit}$ more point sources in the foreground contamination are removed, and thus the S/N increases.
However, only models XRB and Total have S/N$ \ge 1$. As a reference, at $l=[1000,2000]$ and $z=7$ their S/N $\sim 2$ for $\rm X_{limit}=10^{-17}\, \rm erg\, cm^{-2}\, s^{-1}$, while S/N$\sim 1$ for $\rm X_{limit}=10^{-16}\, \rm erg\, cm^{-2}\, s^{-1}$.
It is harder to detect the signal at higher redshifts (at $z=9$ S/N$<1$ in all models) or at larger multipole moments, due to the weaker correlation.

\begin{figure}
\centering
\includegraphics[width=0.9\linewidth]{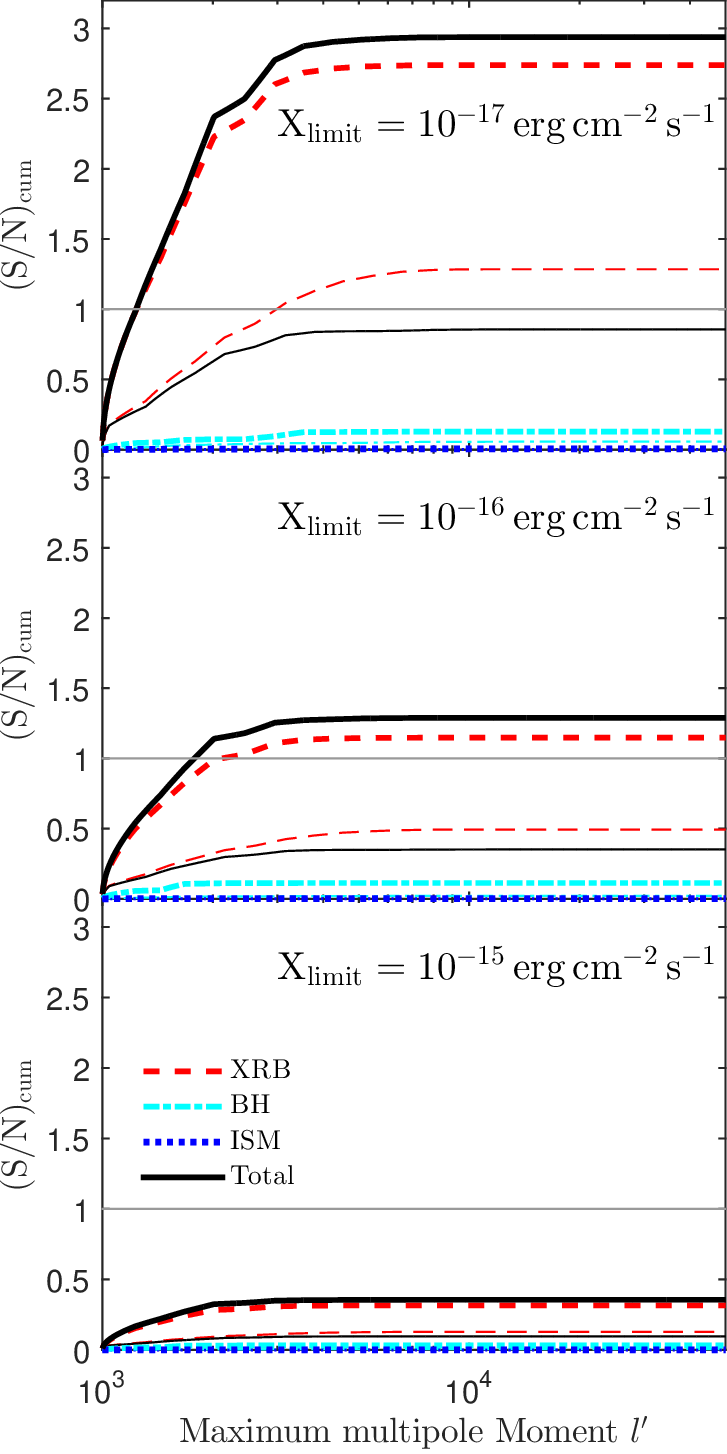}
\caption{Cumulative S/N as a function of the maximum multipole moment $l'$ with X-ray flux limits of $10^{-17}$ (top panel), $10^{-16}$ (center) and $10^{-15}$ (bottom) $\, \rm erg\, cm^{-2}\, s^{-1}$ for model XRB (dashed red lines), BH (dash-dotted cyan), ISM (dotted blue)  and Total (solid black) at $z=7$ (thick lines) and $z=9$ (thin). 
The gray solid horizontal line denotes S/N$=1$.}
\label{sn_cum}
\end{figure}

Fig.~\ref{sn_cum} shows the cumulative S/N (from $l=1000$) in our four models, which is evaluated as:
\begin{equation}
    \left(\frac{S}{N}\right)_{\rm cum}^{2}=\sum_{l=1000}^{l'}\frac{f_{\rm sky}(2l+1) C_{\rm X-21cm}^{2}} {(C_{\rm X}+N_{\rm X}) (C_{\rm 21cm} + N_{\rm 21cm})+C_{\rm X-21cm}^{2}}.
\end{equation}
(S/N)$_{\rm cum}$ is mainly contributed by scales at $l<10^{4}$, consistently with the weak X-ray and 21 cm correlations at $l \ge 10^{4}$ (see Fig.~\ref{ps_xray_21cm_c_vl}).
At $z=7$, the XRB and Total models have (S/N)$_{\rm cum} \sim 3$ at $l'\ge 10^{4}$ in the case of $\rm X_{limit}=10^{-17}\, \rm erg\, cm^{-2}\, s^{-1}$, while at $z=9$ only the XRB model has (S/N)$_{\rm cum} >1$.
With $\rm X_{limit}=10^{-16}\, \rm erg\, cm^{-2}\, s^{-1}$,  (S/N)$_{\rm cum}\sim 1.2$ and 0.5 at $z=7$ and 9.

\begin{figure}
\centering
\includegraphics[width=0.9\linewidth]{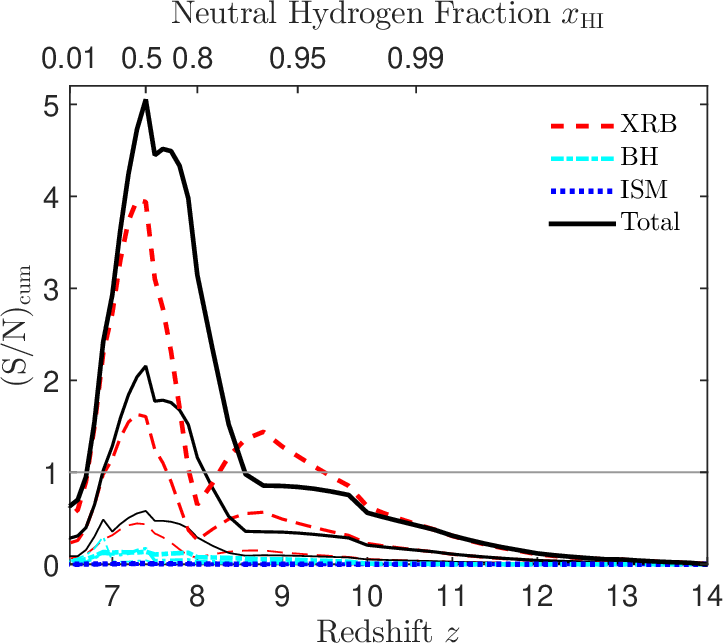}
\caption{Cumulative S/N from $l=1000$ to $10^{4}$ as a function of redshift (or hydrogen neutral fraction) for X-ray flux limits $10^{-17}$ (thick lines), $10^{-16}$ (normal) and $10^{-15}$ (thin) $\, \rm erg\, cm^{-2}\, s^{-1}$ in the model XRB (dashed red lines), BH (dash-dotted cyan), ISM (dotted blue)  and Total (solid black).
The gray solid horizontal line denotes S/N$=1$. 
}
\label{sn_bin_vz}
\end{figure}
Fig.~\ref{sn_bin_vz} shows the evolution of the cumulative S/N from $l=1000$ to $10^{4}$ (i.e. $l'=10^{4}$) in our four source models with different X-ray flux limits. 
For $\rm X_{limit}=10^{-17}\, \rm erg\, cm^{-2}\, s^{-1}$, the correlations in the XRB and Total models can be measured in a wide redshift range from $z=6.7$ to $\sim 9$, with a peak (S/N)$_{\rm cum} \sim 5$ at $z\sim7.5$ where $x_{\rm HI}=0.5$. 
For $\rm X_{limit}=10^{-16}\, \rm erg\, cm^{-2}\, s^{-1}$, (S/N)$_{\rm cum}\ge 1$ at $7<z<7.6$ in the XRB model and at $7<z<8.1$ in the Total model.
It is impossible to measure the correlations for $\rm X_{limit}=10^{-15}\, \rm erg\, cm^{-2}\, s^{-1}$ as predicted by our four models.

We note here that the very deep {\tt Chandra} survey, with a flux limit of $7\times10^{-17}\, \rm erg\, cm^{-2}\, s^{-1}$, 
covers only an area of $0.1 \rm\,deg^{2}$ \citep{Cappelluti2013}, while $9 \rm\,deg^{2}$ are covered by a survey with a flux limit of only $2\times 10^{-15}\, \rm erg\, cm^{-2}\, s^{-1}$ \citep{Kolodzig2017}. Then, such X-ray surveys are not expected to be able to measure the X-ray and 21 cm correlation.
A much larger area of $140 \rm\,deg^{2}$ is expected to be covered by {\tt eROSITA}\footnote{http://www.mpe.mpg.de/eROSITA}, but also in this case sources fainter than $2.9 \times 10^{-15}\, \rm erg\, cm^{-2}\, s^{-1}$ would not be resolved \citep{Merloni2012}.
More promising is the {\tt Athena}\footnote{http://www.the-athena-x-ray-observatory.eu} mission, with a flux limit of $\sim 10^{-16}\, \rm erg\, cm^{-2}\, s^{-1}$ expected for a survey area of $100 \rm\,deg^{2}$ (J. Aird, private communication; \citealt{Aird_etal_2013}).
The proposed {\tt LYNX} telescope\footnote{https://wwwastro.msfc.nasa.gov/lynx} should also be  able to measure such correlation thanks to its designed high resolution.
Deep $\sim$1000~hr integration is considered for {\tt SKA} on five separate $20 \rm\,deg^{2}$ windows covering a total of $100 \rm\,deg^{2}$ \citep{Koopmans2015}.

%
%
\section{Discussion and Conclusions}
\label{sec:dis}

We used the high resolution cosmological hydrodynamical simulation MassiveBlack-II \citep{Khandai2015} post-processed with radiative transfer (RT) calculations (Eide2018a,b) to investigate the high-$z$ component of the cosmic X-ray background (CXB) and its correlation with the 21 cm signal from the epoch of reionization (EoR). We have considered four models with contribution from different source types: X-ray binaries (XRB) \citep{Fragos2013b,Madau2017}, accreting nuclear black holes (BH) \citep{Shakura1973,Krawczyk2013}, hot interstellar medium (ISM) \citep{Mineo2012_ism,Pacucci2014} and a case which includes all the above sources combined (Total). The X-ray sources are treated in the RT simulations together with the UV radiation from stars. 

We found that the global X-ray flux increases rapidly with decreasing redshift, due to the fast growing of star formation. 
The global flux in both the soft (0.5-2)~keV and hard (2-8)~keV X-ray bands at $z>5$ is dominated by the XRBs ($\sim 80\%$ of the Total model), while the BHs and the ISM contribute only $\sim 20\%$ and $<1\%$, respectively.

As BHs have a very low number density but a high luminosity, they display a shot noise like power spectrum, with an amplitude much higher than that of XRBs and ISM.
However, if bright sources could be identified in deep X-ray surveys and removed, the shot noise spectrum of BHs would be strongly reduced \citep{Helgason2014}. The ISM contribution is always negligible. 

We found that the correlation between the CXB and the 21 cm signal is significant at $l<10^{4}$, while it is almost zero at larger $l$. 
The correlations are positive at high $z$ when most of the gas is in a neutral state and the overdensity distribution dominates the signal, while they become negative once most of the hydrogen becomes ionized. 
The transition from a positive to a negative correlation depends on both the X-ray model and the angular scale considered, i.e. it happens earlier in models in which the X-ray flux is stronger and on scales ionized earlier (the smaller scales). As a reference, the transition in the Total model happens at $z=9.1$ and 8.2 for $l=5000$ and $1000$, respectively, while in the XRB model the transition at the same scales happens at $z=8.3$ and 7.8, respectively.

The detectability of the X-ray and 21 cm correlations is highly sensitive to the resolution of the X-ray surveys, as the noise level expected for the deep surveys planned by {\tt SKA} is much smaller than the power spectrum and thus is not expected to affect the measurements. 
We found that if the X-ray survey is deep enough to remove the bright sources with an observed flux $>10^{-17}\, \rm erg\, cm^{-2}\, s^{-1}$, the cumulative signal-to-noise ratio (S/N)$_{\rm cum}$ from $l=1000$ to $10^{4}$ would be $\sim 5$ at $x_{\rm HI}=0.5$, and $\sim 2$ if the sources with observed flux $>10^{-16}\, \rm erg\, cm^{-2}\, s^{-1}$ are removed, while if only sources with observed flux $>10^{-15}\, \rm erg\, cm^{-2}\, s^{-1}$ are removed, the correlations could not be measured. 
It will also be crucial to cover a large enough survey area to obtain a high S/N, although this requires longer total exposure time for both the X-ray and 21~cm facilities.
Such large area surveys would allow to measure the correlations also at scales of $l<1000$ (not covered by our simulations), where the X-ray background and 21 cm signals are still expected to correlate \citep{Liang2016}. 

Although different models and approaches are adopted, our conclusions have a broad agreement to those in \cite{Liang2016}, especially with regard to the specific behaviour of the evolution of X-ray background and the 21 cm signal correlations.

Our predicted CXB from the EoR is much lower than what has been observed, both in terms of global X-ray flux (less than a few percent of the one measured by \citealt{Cappelluti2017} and the one inferred by \citealt{Lehmer2012}) and of power spectrum of fluctuations ($\lesssim 2\%$ of those measured by \citealt{Cappelluti2013} and \citealt{Kolodzig2017}). This suggests that the X-ray contribution from high-$z$ energetic sources could be larger than the one considered here, leaving some freedom in the choice of some parameters adopted in the simulation.
In particular, the BHs properties and distribution at these redshifts are strongly dependent on the seeding procedure adopted in the hydrodynamical simulations (see Eide2018a and Eide2018b for a discussion) and a different procedure could easily increase their contribution without violating observational constraints \citep{Salvaterra2007}.

In conclusion, the X-ray radiation and 21 cm signal during the EoR show significant correlations, which could be used in the future to reduce systematic effects in either X-ray or 21 cm data, as well as to confirm the cosmological origin of the 21~cm signal and to help constraining the properties of X-ray sources during the EoR.
The combination of the planned 21 cm experiment {\tt SKA} and X-ray facilities such as {\tt LynX} and {\tt Athena} in the near future have the potential to measure such correlations with a meaningful S/N.

\section*{Acknowledgements}
The authors are grateful to James Aird, Andrea Merloni, Arne Rau and Mara Salvato for inputs on planned X-ray surveys, Ruben Salvaterra for useful comments, and an anonimous referee for his/her comments.
The tools for bibliographic research are offered by the NASA Astrophysics Data Systems and by the JSTOR archive.
QM is supported by the National Basic Research Program (“973” Program) of China (grant No. 2014CB845800), the National Natural Science Foundation of China (grant Nos. 11673068 and 11725314), the Youth Innovation Promotion Association (2011231), the Key Research Program of Frontier Sciences (QYZDB-SSW-SYS005), the Strategic Priority Research Program "Multiwaveband gravitational wave Universe" (grant No. XDB23000000) of the Chinese Academy of Sciences.
MBE thanks the Astronomy and Astrophysics department at UCSC for their kind hospitality and is a fellow of the U.S.-Norway Fulbright Foundation. 
KH acknowledges support from the Icelandic Research Fund, grant number 173728-051.

%
\appendix

\bibliographystyle{mnras}
\bibliography{ref}

\bsp	
\label{lastpage}
\end{document}